\documentstyle[twoside]{p}

\begin{document}

\title{NONLINEAR DYNAMICS IN ONE AND TWO DIMENSIONAL\\ ARRAYS OF
DISCRETE JOSEPHSON ELEMENTS}

\author{Robert D. Parmentier}

\address{Dipartimento di Fisica, Universit\`{a} di Salerno,\\ I-84081
Baronissi (SA), Italy}

\abstract{Discrete arrays of Josephson junction elements differ from their
continuum counterparts in two essential ways: i) localized dynamic states
in discrete arrays, which are not present in the corresponding continuum
system, can interact with other excitations that are present; ii) fluxoid
quantization for the non-superconducting `holes' provides a constraint for
discrete arrays that is not present in the corresponding continuum system.
The consequences of these effects in one-dimensional systems are now
beginning to be understood; in two-dimesional systems, on the other hand,
the picture is not yet altogether clear. Progress in fabrication technology
and potential applications in practical electronic devices -- as well as
intrinsic interest in nonlinear dynamics -- have contributed significantly
to the growing interest in these systems.}

\kapitola{I. INTRODUCTION}

Progress in thin film and photolithographic technology has permitted the
construction of large one- and two-dimensional planar arrays of Josephson
tunnel junctions having rather precisely designed characteristics;
simultaneously, the increasing availability of computing power has
permitted large-scale simulations of such arrays, even using small
desk-top machines. Planar Josephson junction arrays have attracted research
interest both because they display a rich variety of complicated nonlinear
behaviors and hence can serve as convenient model systems for studying,
{\em e.g.}, the magnetic behavior of granular superconducting materials,
properties of phase transitions in low-dimensional systems, the interplay
between coherence, chaos, pattern formation {\em etc.}, in complex systems,
and also because of current or potential applications in practical
electronic devices, {\em e.g.}, voltage standards, logic circuits, and
millimeter-wave oscillators and amplifiers for radio astronomy and
space-borne receivers. Although much can be intuited about the behavior of
discrete planar arrays from the behavior of the corresponding continuum
systems, discreteness introduces a number of aspects into the dynamics of
such arrays that have no counterpart in the continuum systems, a prime
example being localized dynamic states.

\kapitola{II. ONE-DIMENSIONAL ARRAYS}

A first step in the direction of understanding the dynamics of a
one-dimensional {\it discrete} array may be obtained by considering the
dynamics of the corresponding one-dimensional {\it continuum} structure.
For a one-dimensional, overlap-geometry \cite{Pedersen86} continuum
junction, the partial differential equation (PDE) that describes the
dynamics of the system, in normalized form, is
\begin{equation} 
\phi_{xx} - \phi_{tt} - \sin \phi = \alpha \phi_{t} - \beta \phi_{xxt} -
\gamma \, ,
\end{equation}
with the boundary conditions
\begin{equation} 
\phi_{x}(0,t) + \beta \phi_{xt}(0,t) = \phi_{x}(L,t) + \beta
\phi_{xt}(L,t) = \eta \, .
\end{equation}
In Eq.~(1), $\phi$ is the quantum phase difference between the two
superconducting electrodes of the junction, $\phi_{t}$ is the voltage,
$\alpha$ is a dissipative term due to quasi-particle tunneling (normally
assumed ohmic), $\beta$ is a dissipative term due to surface resistance of
the superconductors, $\gamma$ is a normalized bias current, and $x$ and $t$
are normalized space and time, respectively. In Eq.~(2), $\eta$ is a
normalized magnetic field applied in the plane of the junction,
perpendicular to its long dimension, and $L$ is the normalized junction
length.

Perhaps the most significant feature of the system described by Eqs.~(1-2)
is the propagation of solitons -- called ``fluxons'' in this context
because they carry one quantum of magnetic flux -- in a number of different
dynamic configurations \cite{RDP93}; of particular interest are those
configurations that give rise to the structures in the current-voltage
characteristic of the junction known as zero-field steps, Fiske steps, and
flux-flow steps \cite{RDP93}.

A fairly standard approach for the numerical integration of Eqs.~(1-2)
involves spatial discretization, {\em i.e.}, replacing the spatial
derivatives with finite-difference approximations. For example, replacing
the first spatial derivative with a central-difference approximation and
the second spatial derivative with a three-point approximation \cite{A&S},
and assuming an array of $N$ points having a lattice spacing of $a$, yields
directly (we note in passing that different approximations for the
derivatives may change the form of these equations -- especially Eqs.~(3)
and (5) -- slightly)

\medskip

\noindent at point 1:
\begin{equation} 
\frac{dV_{1}}{dt} = \frac{2}{a^{2}} (\phi_{2} - \phi_{1}) + \frac{2
\beta}{a^{2}} (V_{2} - V_{1}) - \sin \phi_{1} - \alpha V_{1} + \gamma -
\frac{2\eta}{a} \, ;
\end{equation}

\medskip

\noindent at point $n$, $2 \leq n \leq N-1$:
\begin{equation} 
\frac{dV_{n}}{dt} = \frac{1}{a^{2}} (\phi_{n-1} - 2\phi_{n} + \phi_{n+1})
+ \frac{\beta}{a^{2}} (V_{n-1} - 2V_{n} + V_{n+1}) - \sin \phi_{n} - \alpha
V_{n} + \gamma \, ;
\end{equation}

\medskip

\noindent at point $N$:
\begin{equation} 
\frac{dV_{N}}{dt} = \frac{2}{a^{2}} (\phi_{N-1} - \phi_{N}) +
\frac{2\beta}{a^{2}} (V_{N-1} - V_{N}) - \sin \phi_{N} - \alpha V_{N} +
\gamma + \frac{2\eta}{a} \, ;
\end{equation}

\medskip

\noindent at all points, $1 \leq n \leq N$:
\begin{equation} 
\frac{d\phi_{n}}{dt} = V_{n} \, .
\end{equation}
In this way we have obtained $2N$ first-order ordinary differential
equations (ODEs) in $2N$ time-dependent variables ($N$ phases and $N$
voltages), which are just the Kirchhoff circuit-law equations for an array
of $N$ discrete Josephson junction elements interconnected via a parallel
resistance/inductance combination.

Since Eqs.~(3-6) have been obtained as an approximation to Eqs.~(1-2), it
is reasonable to ask to what extent the solutions of the ODE system will be
a reasonable approximation to the solutions of the PDE system. This
question was explored numerically some years ago by Currie {\em et al.}
\cite{Currie77}, who showed that discreteness effects are small as long as
$a \ll 1$, but that as $a \rightarrow 1$, they become appreciable.
Following through the normalizations employed in Eqs.~(1-6), it turns out
that for $a \sim 1$, the `holes' in the discrete array are large enough
to contain $\sim 1$ flux quantum, $h/2e$ ($h$ is Planck's constant and
$e$ the electron charge). Thus, a fluxon propagating through a discrete
array having $a \ll 1$ feels the discreteness only as a slight `bumpiness
in the road', whereas with $a \sim 1$ it can become trapped in the
potential well that is formed between adjacent junctions, in the sense
that it must acquire a certain minimum energy, the so-called
Peierls-Nabarro barrier energy, in order to proceed.

The numerical work of Currie {\em et al.} \cite{Currie77} showed that the
shape of a fluxon propagating through a discrete array is modulated by the
discreteness, thus giving rise to the generation of small-amplitude
oscillations. This mechanism was elucidated by Peyrard and Kruskal
\cite{Peyrard84}, who proposed an analysis based on linearizing the
equations for these small oscillations and seeking stationary solutions of
the linearized equations. Their analysis succeeded remarkably in accounting
for many of the salient features of the observed numerical solutions of the
full equations, in particular, the fact that a fluxon propagating at
certain well-defined velocities generates very little radiation and thus
propagates in a quasi-stationary manner, whereas at other velocities the
radiation of small-amplitude oscillations is large, causing a rapid
deceleration of the fluxon. The existence of special velocities for
quasi-stationary propagation was given further numerical underpinning by
the work of Duncan {\em et al.} \cite{Duncan93}, who also gave further
confirmation to the suggestion of Peyrard and Kruskal \cite{Peyrard84} that
this phenomenon is not peculiar to Josephson junction arrays, but is
present in many nonlinear lattice systems.

The idea of linearizing the equation for the small oscillations radiated
by a modulated fluxon was developed further by Ustinov {\em et al.}
\cite{Ustinov93}. Their point of departure was the observation that the
dispersion relation for small-amplitude linear waves in a discrete array
is qualitatively different from that in the corresponding continuum
system \cite{Brillouin}. For example, for Eq.~(1) (with dissipative and
energy-input terms set to zero and assuming an infinite-length system), the
dispersion relation is
\begin{equation} 
\omega^{2} = 1 + k^{2} \, ,
\end{equation}
where $\omega$ is the angular frequency and $k$ is the wave number. For
Eq.~(4), instead, the dispersion relation is
\begin{equation} 
\omega^{2} = 1 + \frac{4}{a^{2}} \sin^{2}\left(\frac{ka}{2}\right) \, ,
\end{equation}
which coincides with Eq.~(7) only in the limit $a \rightarrow 0$. From
Eq.~(8), Ustinov {\em et al.} \cite{Ustinov93} calculated the phase
velocity, $\omega/k$, for small-amplitude oscillations and found conditions
for resonant interactions with a propagating fluxon. Such superradiant,
{\em i.e.}, phase-locked, interactions give rise to a series of sub-steps
in the zero-field step that would be present in the case of simple fluxon
propagation. Although the analysis of Ustinov {\em et al.} \cite{Ustinov93}
was performed for an annular-geometry array, {\em i.e.}, an array with
periodic boundary conditions in place of the finite boundary conditions of
Eqs.~(3) and (5), studies by Costabile and Sabatino and by Rotoli
\cite{Costabile93} showed that similar effects are present in arrays with
finite boundary conditions, even though the phenomenon is rendered somewhat
more complicated by the effects of reflections from the ends of the array.

As mentioned in the Introduction, much of the impetus for the study of
Josephson junction arrays has come from various practical electronic
applications. Although I cannot here delve deeply into this topic, perhaps
it would be appropriate to mention briefly a few a these.

One of the most firmly established applications is that of the Josephson
voltage standard, which is based on the fact that the relation between the
frequency of a microwave signal applied to a Josephson junction and the
resulting voltage that appears across its electrodes depends only on the
fundamental quantity, $2e/h$; this provides the possibility of generating
voltages that are known to an extremely high degree of precision. However,
the voltages that can be obtained using a single junction are on the order
of several millivolts, rather lower than the values of 1--10 V which would
be most convenient for laboratory use. For this reason, people have for a
number of years moved in the direction of employing series-biased arrays of
thousands of junctions, an idea which is practicable if and only if all of
the individual junctions in the array can be coherently phase locked to a
single microwave source. The state of the art in this area has recently
been reviewed by Niemeyer \cite{Niemeyer92}.

The Rapid Single Flux Quantum (RSFQ) family of logic devices has begun to
attract a growing level of research attention in recent years because it
offers substantial promise for the realization of computing and other
digital signal processing circuits operating at hundreds of GHz, at
extremely low power-dissipation levels, and with comfortably wide
parameter-margin tolerances. One of the basic elements of RSFQ circuits is
the one-dimensional array of small Josephson junctions used essentially as
a transmission line \cite{Seva94}. Such arrays provide a convenient means
for transfering SFQ pulses between active elements, for amplifying the
magnetic field energy connected with a flux quantum, for providing
calibrated time delay, and, with appropriate bias currents, for
generating and injecting a train of SFQ into a circuit. Likharev
\cite{Likharev93} has recently reviewed the most recent achievements in
this area.

High-frequency amplifiers based on Josephson arrays have also been
studied outside of the RSFQ context. Particular attention has been
dedicated to arrays of bridge-type Josephson elements constructed by
appropriately patterning thin films of high-T$_{c}$ superconductors. An
indicative example of the performance features obtainable using this
technology is the SFFT (superconducting flux-flow transistor) amplifier
described by Martens {\em et al.} \cite{Martens93} at the 1992 Applied
Superconductivity Conference: this device showed a gain of 7 dB over a
bandwidth of 50 GHz.

Millimeter-wave oscillators using Josephson arrays -- both one- and
two-dimen-\linebreak sional -- continue to attract active research
interest. A significant stimulus for the study of Josephson millimeter-wave
amplifiers and oscillators is undoubtedly the fact that another Josephson
element, the SIS (superconductor-insulator-superconductor) mixer
\cite{Blundell91}, is already firmly established as the best choice as a
low-noise front-end detector in the range from $\sim$~100~GHz to
$\sim$~1~THz, since its intrinsic noise temperature seems to limited only
by fundamental quantum-uncertainty effects. Consequently, the idea of a
fully integrated superconducting receiver assumes considerable importance,
especially for space-borne communications and radio-astronomical systems in
which high sensitivity and low weight and volume are crucial. The state of
the art in the area of Josephson-array millimeter-wave oscillators was
reviewed a few years ago by Lukens \cite{Lukens90} and up-dated recently by
Bi {\em et al.} \cite{Lukens93}.

\kapitola{III. TWO-DIMENSIONAL ARRAYS}

The passage from a two-dimensional continuum system to the corresponding
discrete array is somewhat less transparent than the passage from
Eqs.~(1-2) to Eqs.~(3-6): in addition to the obvious choice of a square (or
rectangular) array, one can also well imagine two-dimensional arrays
having, {\em e.g.}, a triangular or hexagonal unit cell
\cite{Zagrodzinski93}. Moreover, two-dimensional continuum systems have
only recently (see, {\em e.g.}, \cite{Lachenmann93}) begun to receive the
detailed attention that has for years been dedicated one-dimensional
systems, so that acquired intuition provides us less help in this case.

Whatever the form of the unit cell, the condition of fluxoid quantization
\cite{Bar/Pat} imposes that the relation between the sum of the phase
differences across the junctions in a closed loop and the fluxoid $\Phi$
traversing the loop is given by
\begin{equation} 
\sum_{loop} \phi = -2\pi \frac{\Phi}{\Phi_{0}} + 2\pi n \, ,
\end{equation}
where $n$ is an integer and $\Phi_{0} = h/2e$ is the flux quantum. The
ratio $\Phi/\Phi_{0}$ is known as the frustration, $f$. Once the
structure of the unit cell is defined, the Kirchhoff circuit laws can be
written in a fairly straightforward generalization of Eqs.~(3-6). These,
together with Eq.~(9), completely define the dynamics of the model.

The fluxoid $\Phi$ may be divided into two components, one due to an
externally-applied magnetic field perpendicular to the array and the
other due to self-induced fields stemming from both the self inductance of
a given loop and the mutual inductance between the given loop and its
neighbors, {\em i.e.}
\begin{equation} 
\Phi = \Phi_{ext} + \Phi_{ind} \, .
\end{equation}
There is not yet a universal agreement in the literature regarding the
appropriate form to use for the mutual-inductance contribution to
$\Phi_{ind}$, even in the simplest case of a square array. There is some
indication \cite{Lomdahl93} that when $f \neq 0$ it may be a reasonable
approximation simply to neglect this contribution; on the other hand, at
least when $\Phi_{ext} = 0$, it appears important to take into account at
least nearest-neighbor mutual-inductance terms \cite{Dominguez92}. The
question certainly requires and deserves further study.

Although interest in two-dimensional arrays in fact dates back a number
of years \cite{Lobb83} in connection with the analogy to the frustrated
$XY$-model in spin glass theory, many of the current experimental and
numerical studies of such arrays have been dedicated to understanding the
dynamics associated with the fractional giant Shapiro steps first reported
by Benz {\em et al.} \cite{Benz90}: constant-voltage steps in the dc
current-voltage characteristic of dc + ac current-driven arrays, in the
presence of frustration $f = p/q$, with $p$ and $q$ relative primes, at
average-voltage values given by
\begin{equation} 
V_{q} = \frac{m}{q} M \Phi_{0} \nu \, ,
\end{equation}
where $\nu$ is the frequency of the ac-current drive, $M$ is the number of
junctions in the array along the current direction (an $M \times N$
rectangular array is assumed), and $m$ and $q$ are integers. Detailed
numerical simulations together with mechanical analog studies
\cite{Miguel94} suggest that different dynamic mechanisms might be present
in different regions of parameter space. The question may possibly be
further elucidated by using the powerful technique of low temperature
scanning electron microscopy (LTSEM), which has already shown promising
results for dc-driven arrays \cite{Lachenmann94}. Undoubtedly as these
studies progress, more complicated dynamical states, analogous, {\em e.g.},
to the chaotic states already observed in one-dimensional arrays
\cite{Dominguez93}, will be uncovered.

One of the `applications' of two dimensional arrays that has stimulated
research activity in recent times is the use of such an array as a model
for a sample of high-T$_{c}$ superconductor. Samples of these materials
frequently have a granular structure; one can consider such a sample to be
a network of superconducting grains mutually coupled to their nearest
neighbors via Josephson junctions at their points of contact and containing
non-superconducting intergranular regions. A two-dimensional Josephson
array can be considered a reasonable model for a thin film of high-T$_{c}$
granular material in a perpendicular magnetic field or for a cylindrically
symmetric sample in a uniform axial field. Study of such arrays has
provided useful insight into the mechanism of low-field magnetic
penetration into high-T$_{c}$ samples \cite{Pace93}. For these studies, a
correct treatment of the mutual-inductance terms in the array assumes an
important r\^{o}le \cite{Pace91}.

\kapitola{IV. CONCLUSIONS}

Research on one- and two-dimensional arrays of discrete Josephson elements
is proceeding apace, with new results emerging at a truly surprising rate.
Fundamental studies of nonlinear dynamics are proceeding hand in hand with
practical electronic applications. So far, one-dimensional arrays are
undoubtedly more studied and better understood than their two-dimensional
cousins. Numerical simulation studies of two-dimensional arrays have
outnumbered experimental measurements, and these experimental studies
have been largely limited to systems based on overdamped, SNS-type
junctions, but the picture has already begun to change. I predict that this
development will continue for some time as researchers gradually acquire
more powerful technological and computational tools as well as a more solid
physical intuition regarding the complicated behavior of these systems.

Finally, I would like to mention in passing that -- once again
stimulated, or at least facilitated, by progress in fabrication
technology -- in addition to {\it planar} arrays, the topic of {\it
vertically-stacked} junction arrays has recently begun to attract an
increasing amount of research attention \cite{Hermann94}. Also in this
case the use of such structures as model systems, in this case for
high-T$_{c}$ layered cuprate materials \cite{Paul94}, together with
potential practical electronic applications \cite{Gundlach93}, have
provided much of the essential motivation.

\bigskip

\centerline{\bf Acknowledgements}

It is a pleasure to thank Miguel Octavio for a critical reading of the
manuscript. Financial support from the European Union under contract no.
SC1-CT91-0760 (TSTS) of the ``Science'' program, from MURST (Italy), and
from the Progetto Finalizzato ``Tecnologie Superconduttive e Criogeniche''
del CNR (Italy) is gratefully acknowledged.

\end{document}